\documentstyle[12pt]{article}

\newcommand{\comment}[1]{}
\newcommand{\keywords}[1]{}

\begin{document}

\title{The Thermodynamics of Black Holes}

\author{Robert M. Wald\\
         Enrico Fermi Institute and Department of Physics\\
         University of Chicago\\
         5640 S. Ellis Avenue\\
         Chicago, Illinois 60637-1433\\
         email:rmwa@midway.uchicago.edu}
\date{}
\maketitle

\begin{abstract}

We review the present status of black hole thermodynamics. Our
review includes discussion of classical black hole thermodynamics,
Hawking radiation from black holes, the generalized second law, and
the issue of entropy bounds. A brief survey also is given of
approaches to the calculation of black hole entropy. We conclude with
a discussion of some unresolved open issues.

\end{abstract}


\newpage

\section{Introduction}
\label{section:intro}

During the past 30 years, research in the theory of black holes in
general relativity has brought to light strong hints of a very deep
and fundamental relationship between gravitation, thermodynamics, and
quantum theory. The cornerstone of this relationship is black hole
thermodynamics, where it appears that certain laws of black hole
mechanics are, in fact, simply the ordinary laws of thermodynamics
applied to a system containing a black hole. Indeed, the discovery of
the thermodynamic behavior of black holes---achieved primarily by
classical and semiclassical analyses---has given rise to most of our
present physical insights into the nature of quantum phenomena
occurring in strong gravitational fields.

The purpose of this article is to provide a review of the following
aspects of black hole thermodynamics:

\begin{itemize}
\item At the purely classical level, black holes in general relativity
(as well as in other diffeomorphism covariant theories of gravity)
obey certain laws which bear a remarkable mathematical resemblance to
the ordinary laws of thermodynamics. The derivation of these laws of
classical black hole mechanics is reviewed in section \ref{cbht}.

\item Classically, black holes are perfect absorbers but do not emit
anything; their physical temperature is absolute zero. However, in
quantum theory black holes emit Hawking radiation with a perfect
thermal spectrum. This allows a consistent interpretation of the laws of
black hole mechanics as physically corresponding to the ordinary laws
of thermodynamcs. The status of the derivation of Hawking radiation
is reviewed in section \ref{hr}.

\item The {\it generalized second law} (GSL) directly links the laws
of black hole mechanics to the ordinary laws of thermodynamics. The
arguments in favor of the GSL are reviewed in section \ref{gsl}. A
discussion of entropy bounds is also included in this section.

\item The classical laws of black hole mechanics together with the
formula for the temperature of Hawking radiation allow one to identify
a quantity associated with black holes---namely $A/4$ in general
relativity--- as playing the mathematical role of entropy. The apparent
validity of the GSL provides strong evidence that this quantity truly
is the physical entropy of a black hole. A major goal of research in
quantum gravity is to provide an explanation for---and direct
derivation of---the formula for the entropy of a black hole. A brief
survey of work along these lines is provided in section \ref{bhe}.

\item Although much progress has been made in in our understanding of
black hole thermodynamics, many important issues remain
unresolved. Primary among these are the ``black hole information
paradox'' and issues related to the degrees of freedom responsible for
the entropy of a black hole. These unresolved issues are
briefly discussed in section \ref{oi}.

\end{itemize}

Throughout this article, we shall set $G = \hbar = c = k = 1$, and we
shall follow the sign and notational conventions of
\cite{w3}. Although I have attempted to make this review be reasonably
comprehensive and balanced, it should be understood that my choices of
topics and emphasis naturally reflect my own personal viewpoints,
expertise, and biases.

\newpage
\section{Classical Black Hole Thermodynamics}
\label{cbht}

In this section, I will give a brief review of the laws of classical
black hole mechanics.

In physical terms, a black hole is a region where gravity is so strong
that nothing can escape. In order to make this notion precise, one
must have in mind a region of spacetime to which one can contemplate
escaping. For an asymptotically flat spacetime $(M, g_{ab})$
(representing an isolated system), the asymptotic portion of the
spacetime ``near infinity'' is such a region. The {\em black hole}
region, ${\cal B}$, of an asymptotically flat spacetime, $(M,
g_{ab})$, is defined as
\begin{equation}
{\cal B} \equiv M - I^-({\cal I}^+) ,
\label{bh}
\end{equation}
where ${\cal I}^+$ denotes future null infinity and $I^-$ denotes the
chronological past. Similar definitions of a black hole can be given
in other contexts (such as asymptotically anti-deSitter spacetimes)
where there is a well defined asymptotic region.

The {\em event horizon}, ${\cal H}$, of a black hole is defined to be
the boundary of ${\cal B}$. Thus, ${\cal H}$ is the boundary of the
past of ${\cal I}^+$. Consequently, ${\cal H}$ automatically satisfies
all of the properties possessed by past boundaries (see, e.g.,
\cite{he} or \cite{w3} for further discussion).  In particular, ${\cal
H}$ is a null hypersurface\footnote{Since $\cal H$ is a past boundary,
it automatically must be a $C^0$ embedded submanifold (see, e.g.,
\cite{w3}), but it need not be $C^1$. However, essentially all
discussions and analyses of black hole event horizons implicitly
assume $C^1$ or higher order differentiability of $\cal H$. Recently,
this higher order differentiability assumption has been eliminated for
the proof of the area theorem \cite{cdgh}.} which is composed of
future inextendible null geodesics without caustics, i.e., the
expansion, $\theta$, of the null geodesics comprising the horizon
cannot become negatively infinite. Note that the entire future history
of the spacetime must be known before the location of ${\cal H}$ can
be determined, i.e., ${\cal H}$ possesses no distinguished local
significance.

If Einstein's equation holds with matter satisfying the null energy
condition (i.e., if $T_{ab} k^a k^b \geq 0$ for all null $k^a$), then
it follows immediately from the Raychauduri equation (see, e.g.,
\cite{w3}) that if the expansion, $\theta$, of any null geodesic
congruence ever became negative, then $\theta$ would become infinite
within a finite affine parameter, provided, of course, that the
geodesic can be extended that far. If the black hole is {\it strongly
asymptotically predictable}---i.e., if there is a globally hyperbolic
region containing $I^-({\cal I}^+) \cup {\cal H}$---it can be shown
that this implies that $\theta \geq 0$ everywhere on $\cal H$ (see,
e.g., \cite{he}, \cite{w3}). It then follows that the surface area,
$A$, of the event horizon of a black hole can never decrease with
time, as discovered by Hawking \cite{h}.

The area increase law bears a resemblence to the second law of
thermodynamics in that both laws assert that a certain quantity has
the property of never decreasing with time. It might seem that this
resemblence is a very superficial one, since the area law is a theorem
in differential geometry whereas the second law of thermodynamics is
understood to have a statistical origin. Nevertheless, this
resemblence together with the idea that information is irretrievably
lost when a body falls into a black hole led Bekenstein to propose
\cite{b1}, \cite{b2} that a suitable multiple of the area of the event
horizon of a black hole should be interpreted as its entropy, and that
a {\it generalized second law} (GSL) should hold: The sum of the
ordinary entropy of matter outside of a black hole plus a suitable
multiple of the area of a black hole never decreases. We will discuss
this law in detail in section \ref{gsl}.

The remaining laws of thermodynamics deal with equilibrium and
quasi-equilibrium processes. At nearly the same time as Bekenstein
proposed a relationship between the area theorem and the second law of
thermodynamics, Bardeen, Carter, and Hawking \cite{bch} provided a
general proof of certain laws of ``black hole mechanics'' which are
direct mathematical analogs of the zeroth and first laws of
thermodynamics. These laws of black hole mechanics apply to stationary
black holes (although a formulation of these laws in terms of isolated
horizons will be briefly described at the end of this section).

In order to discuss the zeroth and first laws of black hole mechanics,
we must introduce the notions of stationary, static, and axisymmetric
black holes as well as the notion of a Killing horizon. If an
asymptotically flat spacetime $(M, g_{ab})$ contains a black hole,
${\cal B}$, then ${\cal B}$ is said to be {\em stationary} if there
exists a one-parameter group of isometries on $(M, g_{ab})$ generated
by a Killing field $t^a$ which is unit timelike at infinity. The black
hole is said to be {\em static} if it is stationary and if, in
addition, $t^a$ is hypersurface orthogonal. The black hole is said to
be {\em axisymmetric} if there exists a one parameter group of
isometries which correspond to rotations at infinity. A stationary,
axisymmetric black hole is said to possess the ``$t - \phi$
orthogonality property'' if the 2-planes spanned by $t^a$ and the
rotational Killing field $\phi^a$ are orthogonal to a family of
2-dimensional surfaces. The $t - \phi$ orthogonality property holds
for all stationary-axisymmetric black hole solutions to the vacuum
Einstein or Einstein-Maxwell equations (see, e.g., \cite{heu}).

A null surface, ${\cal K}$, whose null generators coincide with the
orbits of a one-parameter group of isometries (so that there is a
Killing field $\xi^a$ normal to ${\cal K}$) is called a {\em Killing
horizon}. There are two independent results (usually referred to as
``rigidity theorems'') that show that in wide variety of cases of
interest, the event horizon, ${\cal H}$, of a stationary black hole
must be a Killing horizon. The first, due to Carter \cite{c}, states
that for a static black hole, the static Killing field $t^a$ must be
normal to the horizon, whereas for a stationary-axisymmetric black
hole with the $t - \phi$ orthogonality property there exists a Killing
field $\xi^a$ of the form
\begin{equation}
\xi^a = t^a + \Omega \phi^a
\label{xi}
\end{equation}
which is normal to the event horizon. The constant $\Omega$ defined by
eq.(\ref{xi}) is called the {\em angular velocity of the horizon}.
Carter's result does not rely on any field equations, but leaves open
the possibility that there could exist stationary black holes without
the above symmetries whose event horizons are not Killing
horizons. The second result, due to Hawking \cite{he} (see also
\cite{frw}), directly proves that in vacuum or electrovac general
relativity, the event horizon of any stationary black hole must be a
Killing horizon. Consequently, if $t^a$ fails to be normal
to the horizon, then there must exist an additional Killing field,
$\xi^a$, which is normal to the horizon, i.e., a stationary black hole
must be nonrotating (from which staticity follows \cite{sw1}, \cite{sw2},
\cite{cw}) or axisymmetric (though not necessarily with the $t - \phi$
orthogonality property). Note that Hawking's theorem makes no
assumptions of symmetries beyond stationarity, but it does rely on the
properties of the field equations of general relativity.

Now, let ${\cal K}$ be any Killing horizon (not necessarily required
to be the event horizon, ${\cal H}$, of a black hole), with normal
Killing field $\xi^a$. Since $\nabla^a (\xi^b \xi_b)$ also is normal
to ${\cal K}$, these vectors must be proportional at every point on
${\cal K}$. Hence, there exists a function, $\kappa$, on ${\cal K}$,
known as the {\em surface gravity} of ${\cal K}$, which is defined by
the equation
\begin{equation}
\nabla^a (\xi^b \xi_b) = -2 \kappa \xi^a
\label{kappa}
\end{equation}
It follows immediately that $\kappa$ must be constant along each null
geodesic generator of ${\cal K}$, but, in general, $\kappa$ can vary
from generator to generator. It is not difficult to show (see, e.g.,
\cite{w3}) that
\begin{equation}
\kappa = {\rm lim} (Va)
\label{sg}
\end{equation}
where $a$ is the magnitude of the acceleration of the orbits of
$\xi^a$ in the region off of $\cal K$ where they are timelike, $V
\equiv (- \xi^a \xi_a)^{1/2}$ is the ``redshift factor'' of $\xi^a$,
and the limit as one approaches ${\cal K}$ is taken. Equation
(\ref{sg}) motivates the terminology ``surface gravity''. Note that
the surface gravity of a black hole is defined only when it is ``in
equilibrium'', i.e., stationary, so that its event horizon is a
Killing horizon. There is no notion of the surface gravity of a
general, non-stationary black hole, although the definition of surface
gravity can be extended to isolated horizons (see below).

In parallel with the two independent ``rigidity theorems'' mentioned
above, there are two independent versions of the zeroth law of black
hole mechanics. The first, due to Carter \cite{c} (see also
\cite{rw2}), states that for any black hole which is static or is
stationary-axisymmetric with the $t - \phi$ orthogonality property,
the surface gravity $\kappa$, must be constant over its event horizon
${\cal H}$. This result is purely geometrical, i.e., it involves no
use of any field equations. The second, due to Bardeen, Carter, and
Hawking \cite{bch} states that if Einstein's equation holds with the
matter stress-energy tensor satisfying the dominant energy condition,
then $\kappa$ must be constant on any Killing horizon. Thus, in the
second version of the zeroth law, the hypothesis that the $t - \phi$
orthogonality property holds is eliminated, but use is made of the
field equations of general relativity.

A {\em bifurcate Killing horizon} is a pair of null surfaces, ${\cal
K}_A$ and ${\cal K}_B$, which intersect on a spacelike 2-surface,
$\cal C$ (called the ``bifurcation surface''), such that ${\cal K}_A$
and ${\cal K}_B$ are each Killing horizons with respect to the same
Killing field $\xi^a$. It follows that $\xi^a$ must vanish on $\cal
C$; conversely, if a Killing field, $\xi^a$, vanishes on a
two-dimensional spacelike surface, ${\cal C}$, then ${\cal C}$ will be
the bifurcation surface of a bifurcate Killing horizon associated with
$\xi^a$ (see \cite{w4} for further discussion). An important
consequence of the zeroth law is that if $\kappa \neq 0$, then in the
``maximally extended'' spacetime representing a stationary black hole,
the event horizon, ${\cal H}$, comprises a branch of a bifurcate
Killing horizon \cite{rw2}. This result is purely
geometrical---involving no use of any field equations. As a
consequence, the study of stationary black holes which satisfy the
zeroth law divides into two cases: ``extremal'' black holes (for
which, by definition, $\kappa = 0$), and black holes with bifurcate
horizons.

The first law of black hole mechanics is simply an identity relating
the changes in mass, $M$, angular momentum, $J$, and horizon area,
$A$, of a stationary black hole when it is perturbed. To
first order, the variations of these quantities in the vacuum case
always satisfy
\begin{equation}
\delta M = \frac{1}{8 \pi} \kappa \delta A + \Omega \delta J.
\label{bh1}
\end{equation}
In the original derivation of this law \cite{bch}, it was required
that the perturbation be stationary. Furthermore, the original
derivation made use of the detailed form of Einstein's
equation. Subsequently, the derivation has been generalized to hold
for non-stationary perturbations \cite{sw1}, \cite{iw}, provided that
the change in area is evaluated at the bifurcation surface, ${\cal
C}$, of the unperturbed black hole\footnote{See \cite{sor1} for a
derivation of the first law for non-stationary perturbations that does
not require evaluation at the bifurcation surface.}. More
significantly, it has been shown \cite{iw} that the validity of this
law depends only on very general properties of the field
equations. Specifically, a version of this law holds for any field
equations derived from a diffeomorphism covariant Lagrangian,
$L$. Such a Lagrangian can always be written in the form
\begin{equation}
L = L \left( g_{ab}; R_{abcd}, \nabla_a R_{bcde},
...;\psi, \nabla_a \psi, ...\right)
\label{lag}
\end{equation}
where $\nabla_a$ denotes the derivative operator associated with
$g_{ab}$, $R_{abcd}$ denotes the Riemann curvature tensor of $g_{ab}$,
and $\psi$ denotes the collection of all matter fields of the theory
(with indices suppressed). An arbitrary (but finite) number of
derivatives of $R_{abcd}$ and $\psi$ are permitted to appear in $L$.
In this more general context, the first law of black hole mechanics is
seen to be a direct consequence of an identity holding for the
variation of the Noether current. The general form of the first law
takes the form
\begin{equation}
\delta M = \frac{\kappa}{2 \pi} \delta S_{\rm bh} + \Omega \delta J + ...,
\label{first}
\end{equation}
where the ``...'' denote possible additional contributions from long
range matter fields, and where
\begin{equation}
S_{\rm bh} \equiv -2 \pi \int_{\cal C} \frac{\delta L}{\delta
R_{abcd}} n_{ab} n_{cd}
\label{Sbh} .
\end{equation}
Here $n_{ab}$ is the binormal to the bifurcation surface $\cal C$
(normalized so that $n_{ab} n^{ab} = -2$), and the functional
derivative is taken by formally viewing the Riemann tensor as a field
which is independent of the metric in eq.(\ref{lag}). For the case of
vacuum general relativity, where $L = R \sqrt{-g}$, a simple
calculation yields
\begin{equation}
S_{\rm bh} = A/4 
\label{Sbh2}
\end{equation}
and eq.(\ref{first}) reduces to eq.(\ref{bh1}).

The close mathematical analogy of the zeroth, first, and second laws
of thermodynamics to corresponding laws of classical black hole
mechanics is broken by the Planck-Nernst form of the third law of
thermodynamics, which states that $S \rightarrow 0$ (or a ``universal
constant'') as $T \rightarrow 0$. The analog of this law fails in
black hole mechanics\footnote{However, analogs of alternative
formulations of the third law do appear to hold for black holes
\cite{i1}.}, since there exist extremal black holes (i.e., black holes
with $\kappa = 0$) with finite $A$. However, there is good reason to
believe that the the ``Planck-Nernst theorem'' should not be viewed as
a fundamental law of thermodynamics \cite{al} but rather as a property
of the density of states near the ground state in the thermodynamic
limit, which happens to be valid for commonly studied
materials. Indeed, examples can be given of ordinary quantum systems
that violate the Planck-Nernst form of the third law in a manner very
similar to the violations of the analog of this law that occur for
black holes \cite{w5}.

As discussed above, the zeroth and first laws of black hole mechanics
have been formulated in the mathematical setting of stationary black
holes whose event horizons are Killing horizons. The requirement of
stationarity applies to the entire spacetime and, indeed, for the
first law, stationarity of the entire spacetime is essential in order
to relate variations of quantities defined at the horizon (like $A$)
to variations of quantities defined at infinity (like $M$ and
$J$). However, it would seem reasonable to expect that the equilibrium
thermodynamic behavior of a black hole would require only a form of
local stationarity at the event horizon. For the formulation of the
first law of black hole mechanics, one would also then need local
definitions of quantities like $M$ and $J$ at the horizon. Such an
approach toward the formulation of the laws of black hole mechanics
has recently been taken via the notion of an {\it isolated horizon},
defined as a null hypersurface with vanishing shear and expansion
satisfying the additional properties stated in \cite{abdf}. (This
definition supercedes the more restrictive definitions given, e.g., in
\cite{abf1}, \cite{abf}, \cite{ac}.) The presence of an isolated
horizon does not require the entire spacetime to be stationary
\cite{lewa}.  A direct analog of the zeroth law for stationary event
horizons can be shown to hold for isolated horizons \cite{afk}. In the
Einstein-Maxwell case, one can demand (via a choice of scaling of the
normal to the isolated horizon as well as a choice of gauge for the
Maxwell field) that the surface gravity and electrostatic potential of
the isolated horizon be functions of only its area and charge. The
requirement that time evolution be symplectic then leads to a version
of the first law of black hole mechanics as well as a (in general,
non-unique) local notion of the energy of the isolated horizon
\cite{afk}.  These results also have been generalized to allow dilaton
couplings \cite{ac} and Yang-Mills fields \cite{cns}, \cite{afk}.

In comparing the laws of black hole mechanics in classical general
relativity with the laws of thermodynamics, it should first be noted
that the black hole uniqueness theorems (see, e.g., \cite{heu})
establish that stationary black holes---i.e., black holes ``in
equilibrium''---are characterized by a small number of parameters,
analogous to the ``state parameters'' of ordinary thermodynamics. In
the corresponding laws, the role of energy, $E$, is played by the
mass, $M$, of the black hole; the role of temperature, $T$, is played
by a constant times the surface gravity, $\kappa$, of the black hole;
and the role of entropy, $S$, is played by a constant times the area,
$A$, of the black hole.  The fact that $E$ and $M$ represent the same
physical quantity provides a strong hint that the mathematical analogy
between the laws of black hole mechanics and the laws of
thermodynamics might be of physical significance. However, as argued
in \cite{bch}, this cannot be the case in classical general
relativity. The physical temperature of a black hole is absolute zero
(see subsection \ref{av} below), so there can be no physical
relationship between $T$ and $\kappa$. Consequently, it also would be
inconsistent to assume a physical relationship between $S$ and $A$. As
we shall now see, this situation changes dramatically when quantum
effects are taken into account.

\newpage
\section{Hawking Radiation}
\label{hr}

In 1974, Hawking \cite{h2} made the startling discovery that the
physical temperature of a black hole is not absolute zero: As a result
of quantum particle creation effects, a black hole radiates to
infinity all species of particles with a perfect black body spectrum,
at temperature (in units with $G=c=\hbar=k=1$)
\begin{equation}
T = \frac{\kappa}{2 \pi} .
\label{T}
\end{equation}
Thus, $\kappa/2 \pi$ truly is the {\em physical} temperature of a
black hole, not merely a quantity playing a role mathematically
analogous to temperature in the laws of black hole mechanics.  In this
section, we review the status of the derivation of the Hawking effect
and also discuss the closely related Unruh effect.

The original derivation of the Hawking effect \cite{h2} made direct
use of the formalism for calculating particle creation in a curved
spacetime that had been developed by Parker \cite{par} and
others. Hawking considered a classical spacetime $(M, g_{ab})$
describing gravitational collapse to a Schwarzschild black hole. He
then considered a free (i.e., linear) quantum field propagating in
this background spacetime, which is initially in its vacuum state
prior to the collapse, and he computed the particle content of the
field at infinity at late times. This calculation involves taking the
positive frequency mode function corresponding to a particle state at
late times, propagating it backwards in time, and determining its
positive and negative frequency parts in the asymptotic past. His
calculation revealed that at late times, the expected number of
particles at infinity corresponds to emission from a perfect black
body (of finite size) at the Hawking temperature, eq.~(\ref{T}). It
should be noted that this result relies only on the analysis of
quantum fields in the region exterior to the black hole, and it does
not make use of any gravitational field equations.

The original Hawking calculation can be straightforwardly generalized
and extended in the following ways. First, one may consider a
spacetime representing an arbitrary gravitational collapse to a black
hole such that the black hole ``settles down'' to a stationary final
state satisfying the zeroth law of black hole mechanics (so that the
surface gravity, $\kappa$, of the black hole final state is constant
over its event horizon). The initial state of the quantum field may be
taken to be any nonsingular state (i.e., any Hadamard state---see,
e.g. \cite{w4}) rather than the initial vacuum state. Finally, it can
be shown \cite{w7} that all aspects of the final state at late times
(i.e., not merely the expected number of particles in each mode)
correspond to black body\footnote{If the black hole is rotating, the
the spectrum seen by an observer at infinity corresponds to what would
emerge from a ``rotating black body''.} thermal radiation emanating
from the black hole at temperature eq.~(\ref{T}).

It should be noted that no infinities arise in the calculation of the
Hawking effect for a free field, so the results are mathematically
well defined, without any need for regularization or
renormalization. The original derivations \cite{h2}, \cite{w7} made
use of notions of ``particles propagating into the black hole'', but
the results for what an observer sees at infinity were shown to be
independent of the ambiguities inherent in such notions and, indeed, a
derivation of the Hawking effect has been given \cite{fh} which
entirely avoids the introduction of any notion of ``particles''.
However, there remains one significant difficultly with the Hawking
derivation: In the calculation of the backward-in-time propagation of
a mode, it is found that the mode undergoes a large blueshift as it
propagates near the event horizon, but there is no correspondingly
large redshift as the mode propagates back through the collapsing
matter into the asymptotic past. Indeed, the net blueshift factor of
the mode is proportional to $\exp(\kappa t)$, where $t$ is the time
that the mode would reach an observer at infinity. Thus, within a time
of order $1/\kappa$ of the formation of a black hole (i.e., $\sim
10^{-5}$ seconds for a one solar mass Schwarzschild black hole), the
Hawking derivation involves (in its intermediate steps) the
propagation of modes of frequency much higher than the Planck
frequency. In this regime, it is difficult to believe in the accuracy
of free field theory---or any other theory known to mankind.

An approach to investigating this issue was first suggested by Unruh
\cite{u2}, who noted that a close analog of the Hawking effect occurs
for quantized sound waves in a fluid undergoing supersonic flow. A
similar blueshifting of the modes quickly brings one into a regime
well outside the domain of validity of the continuum fluid
equations. Unruh suggested replacing the continuum fluid equations
with a more realistic model at high frequencies to see if the fluid
analog of the Hawking effect would still occur. More recently, Unruh
investigated models where the dispersion relation is altered at
ultra-high frequencies, and he found no deviation from the Hawking
prediction \cite{u3}. A variety of alternative models have been
considered by other researchers \cite{bmps}-\cite{jm}. Again,
agreement with the Hawking effect prediction was found in all cases,
despite significant modifications of the theory at high frequencies.

The robustness of the Hawking effect with respect to modifications of
the theory at ultra-high frequency probably can be understood on the
following grounds. One may view the backward-in-time propagation of
modes as consisting of two stages: a first stage where the
blueshifting of the mode brings it into a WKB regime but the
frequencies remain well below the Planck scale, and a second stage
where the continued blueshifting takes one to the Planck scale and
beyond. In the first stage, the usual field theory calculations should
be reliable. On the other hand, after the mode has entered a WKB
regime, it seems plausible that the kinds of modifications to its
propagation laws considered in \cite{u3}-\cite{jm} should not affect
its essential properties, in particular the magnitude of its negative
frequency part.

Indeed, an issue closely related to the validity of the original
Hawking derivation arises if one asks how a uniformly accelerating
observer in Minkowski spacetime perceives the ordinary (inertial)
vacuum state (see below). The outgoing modes of a given frequency
$\omega$ as seen by the accelerating observer at proper time $\tau$
along his worldline correspond to modes of frequency $\sim \omega
\exp(a \tau)$ in a fixed inertial frame. Therefore, at time $\tau \gg
1/a$ one might worry about field-theoretic derivations of what the
accelerating observer would see. However, in this case one can appeal
to Lorentz invariance to argue that what the accelerating observer
sees cannot change with time. It seems likely that one could similarly
argue that the Hawking effect cannot be altered by modifications of
the theory at ultra-high frequencies, provided that these
modifications preserve an appropriate ``local Lorentz invariance'' of
the theory. Thus, there appears to be strong reasons for believing in
the validity of the Hawking effect despite the occurrence of
ultra-high-frequency modes in the derivation.

There is a second, logically independent result---namely, the Unruh
effect \cite{u1} and its generalization to curved spacetime---which
also gives rise to the formula (\ref{T}). Although the Unruh effect is
mathematically very closely related to the Hawking effect, it is
important to distinguish clearly between them.  In its most general
form, the Unruh effect may be stated as follows (see \cite{kw},
\cite{w4} for further discussion): Consider a a classical spacetime
$(M, g_{ab})$ that contains a bifurcate Killing horizon, ${\cal K} =
{\cal K}_A \cup {\cal K}_B$, so that there is a one-parameter group of
isometries whose associated Killing field, $\xi^a$, is normal to
${\cal K}$. Consider a free quantum field on this spacetime. Then
there exists at most one globally nonsingular state of the field which
is invariant under the isometries. Furthermore, in the ``wedges'' of
the spacetime where the isometries have timelike orbits, this state
(if it exists) is a KMS (i.e., thermal equilibrium) state at
temperature (\ref{T}) with respect to the isometries.

Note that in Minkowski spacetime, any one-parameter group of Lorentz
boosts has an associated bifurcate Killing horizon, comprised by two
intersecting null planes. The unique, globally nonsingular state which
is invariant under these isometries is simply the usual (``inertial'')
vacuum state, $|0>$.  In the ``right and left wedges'' of Minkowski
spacetime defined by the Killing horizon, the orbits of the Lorentz
boost isometries are timelike, and, indeed, these orbits correspond to
worldlines of uniformly accelerating observers. If we normalize the
boost Killing field, $b^a$, so that Killing time equals proper time on
an orbit with acceleration $a$, then the surface gravity of the
Killing horizon is $\kappa = a$. An observer following this orbit
would naturally use $b^a$ to define a notion of ``time translation
symmetry''. Consequently, by the above general result, when the field
is in the inertial vacuum state, a uniformly accelerating observer
would describe the field as being in a thermal equilibrium state at
temperature
\begin{equation}
T = \frac{a}{2 \pi} 
\label{Tu}
\end{equation}
as originally discovered by Unruh \cite{u1}. A mathematically rigorous
proof of the Unruh effect in Minkowski spacetime was given by
Bisognano and Wichmann \cite{bw} in work motivated by entirely
different considerations (and done independently of and nearly
simultaneously with the work of Unruh). Furthermore, the
Bisognano-Wichmann theorem is formulated in the general context of
axiomatic quantum field theory, thus establishing that the Unruh
effect is not limited to free field theory.

Although there is a close mathematical relationship between the Unruh
effect and the Hawking effect, it should be emphasized these results
refer to {\em different} states of the quantum field. We can divide
the late time modes of the quantum field in the following manner,
according to the properties that they would have in the analytically
continued spacetime \cite{rw2} representing the asymptotic final
stationary state of the black hole: We refer to modes that would have
emanated from the white hole region of the analytically continued
spacetime as ``UP modes'' and those that would have originated from
infinity as ``IN modes''.  In the Hawking effect, the asymptotic final
state of the quantum field is a state in which the UP modes of the
quantum field are thermally populated at temperature (\ref{T}), but
the IN modes are unpopulated. This state (usually referred to as the
``Unruh vacuum'') would be singular on the white hole horizon in the
analytically continued spacetime. On the other hand, in the Unruh
effect and its generalization to curved spacetimes, the state in
question (usually referred to as the ``Hartle-Hawking vacuum''
\cite{hh}) is globally nonsingular, and {\em all} modes of the quantum
field in the ``left and right wedges'' are thermally
populated.\footnote{The state in which none of the modes in the region
exterior to the black hole are populated is usually referred to as the
``Boulware vacuum''. The Boulware vacuum is singular on both the black
hole and white hole horizons.}

The differences between the Unruh and Hawking effects can be seen
dramatically in the case of a Kerr black hole. For the Kerr black
hole, it can be shown \cite{kw} that there does not exist any globally
nonsingular state of the field which is invariant under the isometries
associated with the Killing horizon, i.e., there does not exist a
``Hartle-Hawking vacuum state'' on Kerr spacetime. However, there is
no difficultly with the derivation of the Hawking effect for Kerr
black holes, i.e., the ``Unruh vacuum state'' does exist.

It should be emphasized that in the Hawking effect, the
temperature (\ref{T}) represents the temperature as measured by an
observer near infinity. For any observer following an orbit of the
Killing field, $\xi^a$, normal to the horizon, the locally measured
temperature of the UP modes is given by
\begin{equation}
T = \frac{\kappa}{2 \pi V} \: ,
\label{ta}
\end{equation}
where $V = (-\xi^a \xi_a)^{1/2}$. In other words, the locally measured
temperature of the Hawking radiation follows the Tolman law. Now, as
one approaches the horizon of the black hole, the UP modes dominate
over the IN modes. Taking eq.(\ref{sg}) into account, we see that $T
\rightarrow a/2 \pi$ as the black hole horizon, $\cal H$, is
approached, i.e., in this limit eq.(\ref{ta}) corresponds to the flat
spacetime Unruh effect.

Equation (\ref{ta}) shows that when quantum effects are taken into
account, a black hole is surrounded by a ``thermal atmosphere'' whose
local temperature as measured by observers following orbits of $\xi^a$
becomes divergent as one approaches the horizon. As we shall see in
the next section, this thermal atmosphere produces important physical
effects on quasi-stationary bodies near the black hole. On the other
hand, it should be emphasized that for a macroscopic black hole,
observers who freely fall into the black hole would not notice any
important quantum effects as they approach and cross the horizon.

\newpage
\section{The Generalized Second Law (GSL)}
\label{gsl}

In this section, we shall review some arguments for the validity of
the generalized second law (GSL). We also shall review the status of
several proposed entropy bounds on matter that have played a role in
discussions and analyses of the GSL.

\subsection{Arguments for the Validity of the GSL}
\label{av}

Even in classical general relativity, there is a serious difficulty
with the ordinary second law of thermodynamics when a black hole is
present, as originally emphasized by J.A. Wheeler: One can simply take
some ordinary matter and drop it into a black hole, where, according
to classical general relativity, it will disappear into a spacetime
singularity. In this process, one loses the entropy initially present
in the matter, and no compensating gain of ordinary entropy occurs, so
the total entropy, $S$, of matter in the universe decreases. One could
attempt to salvage the ordinary second law by invoking the bookkeeping
rule that one must continue to count the entropy of matter dropped
into a black hole as still contributing to the total entropy of the
universe.  However, the second law would then have the status of being
observationally unverifiable.

As already mentioned in section \ref{cbht}, after the area theorem was
proven, Bekenstein \cite{b1}, \cite{b2} proposed a way out of this
difficulty: Assign an entropy, $S_{\rm bh}$, to a black hole given by a
numerical factor of order unity times the area, $A$, of the black hole in
Planck units. Define the {\it generalized entropy}, $S'$, to be the sum of
the ordinary entropy, $S$, of matter outside of a black hole plus the black
hole entropy
\begin{equation}
S' \equiv S + S_{\rm bh}
\label{S'}
\end{equation}
Finally, replace the ordinary second law of thermodynamics by the {\it
generalized second law} (GSL): The total generalized entropy of the
universe never decreases with time.
\begin{equation}
\Delta S' \geq 0 .
\label{GSL}
\end{equation}
Although the ordinary second law will fail when matter is dropped into
a black hole, such a process will tend to increase the area of the
black hole, so there is a possibility that the GSL will hold.

Bekenstein's proposal of the GSL was made prior to the discovery of
Hawking radiation. When Hawking radiation is taken into account, a
serious problem also arises with the second law of black hole
mechanics (i.e., the area theorem): Conservation of energy requires
that an isolated black hole must lose mass in order to compensate for
the energy radiated to infinity by the Hawking process. Indeed, if one
equates the rate of mass loss of the black hole to the energy flux at
infinity due to particle creation, one arrives at the startling
conclusion that an isolated black hole will radiate away all of its
mass within a finite time. During this process of black hole
``evaporation'', $A$ will decrease. Such an area decrease can occur
because the expected stress-energy tensor of quantum matter does not
satisfy the null energy condition---even for matter for which this
condition holds classically---in violation of a key hypothesis of the
area theorem.

However, although the second law of black hole mechanics fails during
the black hole evaporation process, if we adjust the numerical factor
in the definition of $S_{\rm bh}$ to correspond to the identification
of $\kappa/2\pi$ as temperature in the first law of black hole
mechanics---so that, as in eq.~(\ref{Sbh2}) above, we have $S_{\rm bh}
= A/4$ in Planck units---then the GSL continues to hold: Although $A$
decreases, there is at least as much ordinary entropy generated
outside the black hole by the Hawking process. Thus, although the
ordinary second law fails in the presence of black holes and the
second law of black hole mechanics fails when quantum effects are
taken into account, there is a possibility that the GSL may always
hold. If the GSL does hold, it seems clear that we must interpret
$S_{\rm bh}$ as representing the {\it physical} entropy of a black
hole, and that the laws of black hole mechanics must truly represent
the ordinary laws of thermodynamics as applied to black holes. Thus, a
central issue in black hole thermodynamics is whether the GSL holds in
all processes.

It was immediately recognized by Bekenstein \cite{b1} (see also
\cite{bch}) that there is a serious difficulty with the GSL if one
considers a process wherein one carefully lowers a box containing
matter with entropy $S$ and energy $E$ very close to the horizon of a
black hole before dropping it in. Classically, if one could lower the
box arbitrarily close to the horizon before dropping it in, one would
recover all of the energy originally in the box as ``work'' at
infinity. No energy would be delivered to the black hole, so by the
first law of black hole mechanics, eq.~(\ref{first}), the black hole
area, $A$ would not increase. However, one would still get rid of all
of the entropy, $S$, originally in the box, in violation of the
GSL. 

Indeed, this process makes manifest the fact that in classical general
relativity, the physical temperature of a black hole is absolute zero:
The above process is, in effect, a Carnot cycle which converts
``heat'' into ``work'' with $100\%$ efficiency \cite{g}. The
difficulty with the GSL in the above process can be viewed as stemming
from an inconsistency of this fact with the mathematical assignment of
a finite (non-zero) temperature to the black hole required by the
first law of black hole mechanics if one assigns a finite
(non-infinite) entropy to the black hole.

Bekenstein's proposed a resolution of the above difficulty with the
GSL in a quasi-static lowering process by arguing \cite{b1}, \cite{b2}
that it would not be possible to lower a box containing physically
reasonable matter close enough to the horizon of the black hole to
violate the GSL. As will be discussed further in the next sub-section,
this proposed resolution was later refined by postulating a universal
bound on the entropy of systems with a given energy and size
\cite{b3}. However, an alternate resolution was proposed in \cite{uw},
based upon the idea that, when quantum effects are taken into account,
the physical temperature of a black hole is no longer absolute zero,
but rather is the Hawking temperature, $\kappa/2\pi$. Since the
Hawking temperature goes to zero in the limit of a large black hole,
it might appear that quantum effects could not be of much relevance in
this case. However, the despite the fact that Hawking radiation at
infinity is indeed negligible for large black holes, the effects of
the quantum ``thermal atmosphere'' surrounding the black hole are not
negligible on bodies that are quasi-statically lowered toward the
black hole.  The temperature gradient in the thermal atmosphere (see
eq.(\ref{ta})) implies that there is a pressure gradient and,
consequently, a buoyancy force on the box. This buoyancy force becomes
infinitely large in the limit as the box is lowered to the horizon.
As a result of this buoyancy force, the optimal place to drop the box
into the black hole is no longer the horizon but rather the ``floating
point'' of the box, where its weight is equal to the weight of the
displaced thermal atmosphere. The minimum area increase given to the
black hole in the process is no longer zero, but rather turns out
to be an amount just sufficient to prevent any violation of the GSL
from occurring in this process \cite{uw}.

The analysis of \cite{uw} considered only a particular class of
gedankenexperiments for violating the GSL involving the quasi-static
lowering of a box near a black hole. Of course, since one does not
have a general proof of the ordinary second law of
thermodynamics---and, indeed, for finite systems, there should always
be a nonvanishing probability of violating the ordinary second
law---it would not be reasonable to expect to obtain a completely
general proof of the GSL. However, general arguments within the
semiclassical approximation for the validity of the GSL for arbitrary
infinitesimal quasi-static processes have been given in \cite{tz},
\cite{tzp}, and \cite{w4}. These arguments crucially rely on the
presence of the thermal atmosphere surrounding the black hole.
Related arguments for the validity of the GSL have been given in
\cite{fp} and \cite{s}. In \cite{fp}, it is assumed that the incoming
state is a product state of radiation originating from infinity (i.e.,
IN modes) and radiation that would appear to emanate from the white
hole region of the analytically continued spacetime (i.e., UP modes),
and it is argued that the generalized entropy must increase under
unitary evolution. In \cite{s}, it is argued on quite general grounds
that the (generalized) entropy of the state of the region exterior to
the black hole must increase under the assumption that it undergoes
autonomous evolution.

Indeed, it should be noted that if one could violate the GSL for an
infinitesimal quasi-static process in a regime where the black hole
can be treated semi-classically, then it also should be possible to
violate the ordinary second law for a corresponding process involving
a self-gravitating body. Namely, suppose that the GSL could be
violated for an infinitesimal quasi-static process involving, say, a
Schwarzschild black hole of mass $M$ (with $M$ much larger than the
Planck mass). This process might involve lowering matter towards the
black hole and possibly dropping the matter into it. However, an
observer doing this lowering or dropping can ``probe'' only the region
outside of the black hole, so there will be some $r_0 > 2M$ such that
the detailed structure of the black hole will directly enter the
analysis of the process only for $r > r_0$. Now replace the black hole
by a shell of matter of mass $M$ and radius $r_0$, and surround this
shell with a ``{\em real}'' atmosphere of radiation in thermal
equilibrium at the Hawking temperature (\ref{T}) as measured by an
observer at infinity. Then the ordinary second law should be violated
when one performs the same process to the shell surrounded by the
(``real'') thermal atmosphere as one performs to violate the GSL when
the black hole is present. Indeed, the arguments of \cite{tz},
\cite{tzp}, and \cite{w4} do not distinguish between infinitesimal
quasi-static processes involving a black hole as compared with a shell
surrounded by a (``real'') thermal atmosphere at the Hawking
temperature.

In summary, there appear to be strong grounds for believing in the
validity of the GSL.

\subsection{Entropy Bounds}
\label{eb}

As discussed in the previous subsection, for a classical black hole
the GSL would be violated if one could lower a box containing matter
sufficiently close to the black hole before dropping it in. Indeed,
for a Schwarzschild black hole, a simple calculation reveals that if
the size of the box can be neglected, then the GSL would be violated
if one lowered a box containing energy $E$ and entropy $S$ to within a
proper distance $D$ of the bifurcation surface of the event horizon
before dropping it in, where
\begin{equation}
D < S/(2\pi E) .
\label{D}
\end{equation}
(This formula holds independently of the mass, $M$, of the black
hole.)  However, it is far from clear that the finite size of the box
can be neglected if one lowers a box containing physically reasonable
matter this close to the black hole. If it cannot be neglected, then
this proposed counterexample to the GSL would be invalidated.

As already mentioned in the previous subsection, these considerations
led Bekenstein \cite {b3} to propose a universal bound on the
entropy-to-energy ratio of bounded matter, given by\footnote{Here
``$E$'' is normally interpreted as the energy above the ground state;
otherwise, eq.(\ref{SE}) would be trivially violated in cases where
the Casimir energy is negative \cite{page3}---although in such cases in
may still be possible to rescue eq.(\ref{SE}) by postulating a
suitable minimum energy of the box walls \cite{beke}.}
\begin{equation}
S/E \leq 2\pi R
\label{SE}
\end{equation}
where $R$ denotes the ``circumscribing radius'' of the body. Two key
questions one can ask about this bound are: (1) Does it hold in
nature? (2) Is it needed for the validity of the GSL?

With regard to question (1), even in Minkowski spacetime, there exist
many model systems that are physically reasonable (in the sense of
positive energies, causal equations of state, etc.) for which
eq.(\ref{SE}) fails\footnote{For a recent discussion of such
counterexamples to eq.(\ref{SE}), see \cite{page1}, \cite{page2},
\cite{page3}; for counter-arguments to these references, see
\cite{beke}.}. In particular it is easily seen that for a system
consisting of $N$ non-interacting species of particles with identical
properties, eq.(\ref{SE}) must fail when $N$ becomes sufficiently
large. However, for a system of $N$ species of free, massless bosons
or fermions, one must take $N$ to be enormously large \cite{b4} to
violate eq.(\ref{SE}), so it does not appear that nature has chosen to
take advantage of this possible means of violating (\ref{SE}).
Equation (\ref{SE}) also is violated at sufficiently low temperatures
if one defines the entropy, $S$, of the system via the canonical
ensemble, i.e., $S(T) = - {\rm tr} [\rho \ln \rho]$, where $\rho$
denotes the canonical ensemble density matrix, $\rho = \exp(-H/T)/
{\rm tr} [\exp(- H/T)]$, where $H$ is the Hamiltonian. However, a
study of a variety of model systems \cite{b4} indicates that
(\ref{SE}) holds at low temperatures when $S$ is defined via the
microcanonical ensemble, i.e., $S(E) = \ln n$ where $n$ is the density
of quantum states with energy $E$. More generally, eq.(\ref{SE}) has
been shown to hold for a wide variety of systems in flat spacetime
\cite{b4}, \cite{bs}.

The status of eq.(\ref{SE}) in curved spacetime is unclear; indeed,
while there is some ambiguity in how ``$E$'' and ``$R$'' are defined
in Minkowski spacetime \cite{page3}, is very unclear what these
quantities would mean in a general, non-spherically-symmetric
spacetime\footnote{Note that these same difficulties would also plague
attempts to give a mathematically rigorous formulation of the ``hoop
conjecture'' \cite{mtw}.}. With regard to ``$E$'', it has long been
recognized that there is no meaningful local notion of gravitational
energy density in general relativity. Although numerous proposals have
been made to define a notion of ``quasi-local mass'' associated with a
closed $2$-surface (see, e.g., \cite{p1}, \cite{by}), none appear to
have fully satisfactory properties. Although the difficulties with
defining a localized notion of energy are well known, it does not seem
to be as widely recognized that there also are serious difficulties in
defining ``$R$'': Given any spacelike $2$-surface, $\cal C$, in a
$4$-dimensional spacetime and given any open neighborhood, $\cal O$,
of $\cal C$, there exists a spacelike $2$-surface, ${\cal C}'$
(composed of nearly null portions) contained within $\cal O$ with
arbitrarily small area and circumscribing radius. Thus, if one is
given a system confined to a world tube in spacetime, it is far from
clear how to define any notion of the ``externally measured size'' of
the region unless, say, one is given a preferred slicing by spacelike
hypersurfaces. Nevertheless, the fact that eq.(\ref{SE}) holds for the
known black hole solutions (and, indeed, is saturated by the
Schwarzschild black hole) and also plausibly holds for a
self-gravitating spherically symmetric body \cite{swz} provides an
indication that some version of (\ref{SE}) may hold in curved
spacetime.

With regard to question (2), in the previous subsection we reviewed
arguments for the validity of the GSL that did not require the
invocation of any entropy bounds. Thus, the answer to question (2) is
``no'' unless there are deficiencies in the arguments of the previous
section that invalidate their conclusions. A number of such potential
deficiencies have been pointed out by Bekenstein. Specifically, the
analysis and conclusions of \cite{uw} have been criticized by
Bekenstein on the grounds that: (i) A ``thin box'' approximation was
made \cite{b5}. (ii) It is possible to have a box whose {\it contents}
have a greater entropy than unconfined thermal radiation of the same
energy and volume \cite{b5}. (iii) Under certain assumptions
concerning the size/shape of the box, the nature of the thermal
atmosphere, and the location of the floating point, the buoyancy force
of the thermal atmosphere can be shown to be negligible and thus
cannot play a role in enforcing the GSL \cite{b6}. (iv) Under certain
other assumptions, the box size at the floating point will be smaller
than the typical wavelengths in the ambient thermal atmosphere, thus
likely decreasing the magnitude of the buoyancy force \cite{b7}.
Responses to criticism (i) were given in \cite{uw2} and \cite{pw}; a
response to criticism (ii) was given in \cite{uw2}; and
a response to (iii) was given in \cite{pw}. As far as I am a aware, no
response to (iv) has yet been given in the literature except to note
\cite{fmw} that the arguments of \cite{b7} should pose similar
difficulties for the ordinary second law for gedankenexperiments
involving a self-gravitating body (see the end of
subsection \ref{av} above). Thus, my own view is that eq.(\ref{SE})
is not necessary for the validity of the GSL\footnote{It is worth
noting that if the buoyancy effects of the thermal atmosphere were
negligible, the bound (\ref{SE}) also would not be sufficient to
ensure the validity of the GSL for non-spherical bodies: The bound
(\ref{SE}) is formulated in terms of the ``circumscribing radius'',
i.e., the {\it largest} linear dimension, whereas if buoyancy effects
were negligible, the to enforce the GSL one would need a bound of the
form (\ref{SE}) with $R$ being the {\it smallest} linear
dimension.}. However, this conclusion remains controversial; see
\cite{and} for a recent discussion.

More recently, an alternative entropy bound has been proposed: It has
been suggested that the entropy contained within a region whose
boundary has area $A$ must satisfy \cite{th1}, \cite{b8}, \cite{suss}
\begin{equation}
S \leq A/4 .
\label{hb}
\end{equation}
This proposal is closely related to the ``holographic principle'',
which, roughly speaking, states that the physics in any spatial region
can be fully described in terms of the degrees of freedom associated
with the boundary of that region. (The literature on the holographic
principle is far too extensive and rapidly developing to attempt to
give any review of it here.) The bound (\ref{hb}) would follow from
(\ref{SE}) under the additional assumption of small self-gravitation
(so that $E \mathrel{{}^<_\sim} R$). Thus, many of the arguments in
favor of (\ref{SE}) are also applicable to (\ref{hb}). Similarly, the
counterexample to (\ref{SE}) obtained by taking the number, $N$, of
particle species sufficiently large also provides a counterexample to
(\ref{hb}), so it appears that (\ref{hb}) can, in principle, be
violated by physically reasonable systems (although not necessarily by
any systems actually occurring in nature).

Unlike eq.(\ref{SE}), the bound (\ref{hb}) explicitly involves the
gravitational constant $G$ (although we have set $G = 1$ in all of our
formulas), so there is no flat spacetime version of (\ref{hb})
applicable when gravity is ``turned off''. Also unlike (\ref{SE}), the
bound (\ref{hb}) does not make reference to the energy, $E$, contained
within the region, so the difficulty in defining $E$ in curved
spacetime does not affect the formulation of (\ref{hb}). However, the
above difficulty in defining the ``bounding area'', $A$, of a world
tube in a general, curved spacetime remains present (but see below).

The following argument has been given that the bound (\ref{hb}) is
necessary for the validity of the GSL \cite{suss}: Suppose we had a
spherically symmetric system that was not a black hole (so $R > 2 E$)
and which violated the bound (\ref{hb}), so that $S > A/4 = \pi R^2$. Now
collapse a spherical shell of mass $M = R/2 - E$ onto the system. A
Schwarzschild black hole of radius $R$ should result. But the entropy
of such a black hole is $A/4$, so the generalized entropy will
decrease in this process.

I am not aware of any counter-argument in the literature to the
argument given in the previous paragraph, so I will take the
opportunity to give one here. If there were a system which violated
the bound (\ref{hb}), then the above argument shows that it would be
(generalized) entropically unfavorable to collapse that system to a
black hole. I believe that the conclusion one should draw
from this is that, in this circumstance, it should not be possible to
form a black hole. In other words, the bound (\ref{hb}) should be
necessary in order for black holes to be stable or metastable states,
but should not be needed for the validity of the GSL.

This viewpoint is supported by a simple model calculation.  Consider a
massless gas composed of $N$ species of (boson or fermion) gas
particles confined by a spherical box of radius $R$. Then (neglecting
self-gravitational effects and any corrections due to discreteness of
modes) we have
\begin{equation}
S \sim N^{1/4} R^{3/4} E^{3/4}
\label{sgas}
\end{equation}
We wish to consider a configuration that is not already a black hole,
so we need $E < R/2$. To violate (\ref{hb})---and thereby threaten to
violate the GSL by collapsing a shell upon the system---we need to
have $S > \pi R^2$. This means that we need to consider a model with
$N \mathrel{{}^>_\sim} R^2$. For such a model, start with a region $R$
containing matter with $S > \pi R^2$ but with $E < R/2$. If we try to
collapse a shell upon the system to form a black hole of radius $R$,
the collapse time will be $\mathrel{{}^>_\sim} R$. But the Hawking
evaporation timescale in this model is $t_H \sim R^3/N$, since the rate
of Hawking radiation is proportional to $N$. Since $N
\mathrel{{}^>_\sim} R^2$, we have $t_H \mathrel{{}^<_\sim} R$, so the
Hawking evaporation time is shorter than the collapse
time! Consequently, the black hole will never actually form. Rather,
at best it will merely act as a catalyst for converting the original high
entropy confined state into an even higher entropy state of unconfined
Hawking radiation.

As mentioned above, the proposed bound (\ref{hb}) is ill defined in a
general (non-spherically-symmetric) curved spacetime. There also are
other difficulties with (\ref{hb}): In a closed universe, it is not
obvious what constitutes the ``inside'' versus the ``outside'' of the
bounding area. In addition, (\ref{hb}) can be violated near
cosmological and other singularities, where the entropy of suitably
chosen comoving volumes remains bounded away from zero but the area of
the boundary of the region goes to zero.  However, a reformulation of
(\ref{hb}) which is well defined in a general curved spacetime and
which avoids these difficulties has been given by Bousso
\cite{bo1}-\cite{bo3}. Bousso's reformulation can be stated as
follows: Let ${\cal L}$ be a null hypersurface such that the
expansion, $\theta$, of ${\cal L}$ is everywhere non-positive, $\theta
\leq 0$ (or, alternatively, is everywhere non-negative, $\theta \geq
0$). In particular, ${\cal L}$ is not allowed to contain caustics,
where $\theta$ changes sign from $- \infty$ to $+\infty$. Let $B$ be a
spacelike cross-section of $\cal L$. Bousso's reformulation
conjectures that
\begin{equation}
S_{\cal L} \leq A_B/4 .
\label{bb}
\end{equation}
where $A_B$ denotes the area of $B$ and $S_{\cal L}$ denotes the
entropy flux through ${\cal L}$ to the future (or, respectively, the
past) of $B$.

In \cite{fmw} it was argued that the bound (\ref{gbb}) should be valid
in certain ``classical regimes'' (see \cite{fmw}) wherein the local
entropy density of matter is bounded in a suitable manner by the
energy density of matter. Furthermore, the following generalization of
Bousso's bound was proposed: Let ${\cal L}$ be a null hypersurface
which starts at a cross-section, $B$, and terminates at a
cross-section $B'$. Suppose further that $\cal L$ is such that its
expansion, $\theta$, is either everywhere non-negative or everywhere
non-positive. Then
\begin{equation}
S_{\cal L} \leq |A_B - A_{B'}|/4 .
\label{gbb}
\end{equation}

Although we have argued above that the validity of the GSL should not
depend upon the validity of the entropy bounds (\ref{SE}) or
(\ref{hb}), there is a close relationship between the GSL and the
generalized Bousso bound (\ref{gbb}). Namely, as discussed in section
\ref{cbht} above, classically, the event horizon of a black hole is a
null hypersurface satisfying $\theta \geq 0$. Thus, in a classical
regime, the GSL itself would correspond to a special case of the
generalized Bousso bound (\ref{gbb}). This suggests the intriguing
possibility that, in quantum gravity, there might be a more general
formulation of the GSL---perhaps applicable to an arbitrary horizon as
defined on P. 134 of \cite{w4}, not merely to an event horizon of a
black hole---which would reduce to (\ref{gbb}) in a suitable classical
limit.

\newpage
\section{Calculations of Black Hole Entropy}
\label{bhe}

The considerations of the previous sections make a compelling case for
the merger of the laws of black hole mechanics with the laws of
thermodynamics. In particular, they strongly suggest that $S_{\rm bh}$
($= A/4$ in general relativity---see eqs.(\ref{Sbh}) and (\ref{Sbh2})
above) truly represents the physical entropy of a black hole. Now, the
entropy of ordinary matter is understood to arise from the number of
quantum states accessible to the matter at given values of the energy
and other state parameters. One would like to obtain a similar
understanding of why $A/4$ represents the entropy of a black hole in
general relativity by identifying (and counting) the quantum dynamical
degrees of freedom of a black hole. In order to do so, it clearly will
be necessary to go beyond the classical and semiclassical
considerations of the previous sections and consider black holes
within a fully quantum theory of gravity. In this section, we will briefly
summarize some of the main approaches that have been taken to the
direct calculation of the entropy of a black hole.

The first direct quantum calculation of black hole entropy was given
by Gibbons and Hawking \cite{gh} in the context of Euclidean quantum
gravity. They started with a formal, functional integral expression
for the canonical ensemble\footnote{There is an inconsistency in the
use of the canonical ensemble to derive a formula for black hole
entropy, since the entropy of a black hole grows too rapidly with
energy for the canonical ensemble to be defined. (Equivalently, the
heat capacity of a Schwarzschild black hole is negative, so it cannot
come to equilibrium with an infinite heat bath.) A derivation
of black hole entropy using the microcanonical ensemble has been given
in \cite{by2}.} partition function in Euclidean quantum gravity and
evaluated it for a black hole in the ``zero loop'' (i.e, classical)
approximation. As shown in \cite{w6}, the mathematical steps in this
procedure are in direct correspondence with the purely classical
determination of the entropy from the form of the first law of black
hole mechanics. A number of other entropy calculations that have been
given within the formal framework of Euclidean quantum gravity also
can be shown to be equivalent to the classical derivation (see
\cite{iw2} for further discussion). Thus, although the derivation of
\cite{gh} and other related derivations give some intriguing glimpses
into possible deep relationships between black hole thermodynamics and
Euclidean quantum gravity, they do not appear to provide any more
insight than the classical derivation into accounting for the quantum
degrees of freedom that are responsible for black hole entropy.

Another approach to the calculation of black hole entropy has been to
attribute it to the ``entanglement entropy'' resulting from quantum
field correlations between the exterior and interior of the black hole
\cite{bkls}-\cite{hlw}.  As a result of these correlations across the
event horizon, the state of a quantum field when restricted to the
exterior of the black hole is mixed. Indeed, in the absence of a short
distance cutoff, the von Neumann entropy, $- {\rm tr} [{\rho} \ln
{\rho}]$, of any physically reasonable state would diverge. If one now
inserts a short distance cutoff of the order of the Planck scale, one
obtains a von Neumann entropy of the order of the horizon
area\footnote{One might argue that in this approach, the constant of
proportionality between $S_{\rm bh}$ and $A$ should depend upon the
number, $N$, of species of particles, and thus could not equal $1/4$
(independently of $N$). However, it is possible that the
$N$-dependence in the number of states is compensated by an
$N$-dependent renormalization of $G$ \cite{su} and, hence, of the
Planck scale cutoff.}, $A$. Thus, this approach provides a natural way
of accounting for why the entropy of a black hole is proportional to
its surface area. However, the constant of proportionality depends
upon a cutoff and is not (presently) calculable within this
approach. Furthermore, it is far from clear why the black hole horizon
should be singled out for a such special treatment of the quantum
degrees of freedom in its vicinity, since similar quantum field
correlations will exist across any other null surface. Indeed, it is
particularly puzzling why the local degrees of freedom associated with
the horizon should be singled out since, as already noted in section
\ref{cbht} above, the black hole horizon at a given time is defined in
terms of the entire future history of the spacetime and thus has no
distinguished local significance. Finally, since the gravitational
action and field equations play no role in the above derivation, it is
difficult to see how this approach could give rise to a black hole
entropy proportional to eq.(\ref{Sbh}) (rather than proportional to
$A$) in a more general theory of gravity. Similar remarks apply to
approaches which attribute the relevant degrees of freedom to the
``shape'' of the horizon \cite{sork} or to causal links crossing the
horizon \cite{dou}.

A closely related idea has been to attribute the entropy of the black
hole to the ordinary entropy of its thermal atmosphere \cite{th}). If
we assume that the thermal atmosphere behaves like a free, massless
(boson or fermion) gas, its entropy density will be (roughly)
proportional to $T^3$. However, since $T$ diverges near the horizon in
the manner specified by eq.(\ref{ta}), we find that the total entropy
of the thermal atmosphere near the horizon diverges.  This is, in
effect, a new type of ultraviolet catastrophe. It arises because, on
account of arbitrarily large redshifts, there now are infinitely many
modes---of arbitrarily high locally measured frequency---that
contribute a bounded energy as measured at infinity.  To cure this
divergence, it is necessary to impose a cutoff on the locally measured
frequency of the modes. If we impose a cutoff of the order of the
Planck scale, then the thermal atmosphere contributes an entropy of
order the horizon area, $A$, just as in the entanglement entropy
analysis. Indeed, this calculation is really the same as the
entanglement entropy calculation, since the state of a quantum field
outside of the black hole is thermal, so its von Neumann entropy is
equal to its thermodynamic entropy (see also \cite{mu}). Note that the bulk
of the entropy of the thermal atmosphere is highly localized in a
``skin'' surrounding the horizon, whose thickness is of order of the
Planck length.

Since the attribution of black hole entropy to its thermal atmosphere
is essentially equivalent to the entanglement entropy proposal, this
approach has essentially the same strengths and weaknesses as the
entanglement entropy approach. On one hand, it naturally accounts for
a black hole entropy proportional to $A$. On the other hand, this
result depends in an essential way on an uncalculable cutoff, and it
is difficult to see how the analysis could give rise to to
eq.(\ref{Sbh}) in a more general theory of gravity. The preferred
status of the event horizon and the localization of the degrees of
freedom responsible for black hole entropy to a ``Planck length skin''
surrounding the horizon also remain puzzling in this approach.  To see
this more graphically, consider the collapse of a massive spherical
shell of matter. Then, as the shell crosses its Schwarzschild radius,
the spacetime curvature outside of the shell is still negligibly
small.  Nevertheless, within a time of order the Planck time after
the crossing of the Schwarzschild radius, the ``skin'' of thermal
atmosphere surrounding the newly formed black hole will come to
equilibrium with respect to the notion of time translation symmetry
for the static Schwarzschild exterior. Thus, if an entropy is to be
assigned to the thermal atmosphere in the manner suggested by this
proposal, then the degrees of freedom of the thermal
atmosphere---which previously were viewed as irrelevant vacuum
fluctuations making no contribution to entropy---suddenly become
``activated'' by the passage of the shell for the purpose of counting
their entropy. A momentous change in the entropy of matter in the
universe has occurred, even though observers riding on or near the
shell see nothing of significance occurring.

Another approach that is closely related to the entanglement entropy
and thermal atmosphere approaches---and which also contains elements
closely related to the Euclidean approach and the classical derivation
of eq.(\ref{Sbh})---attempts to account for black hole entropy in the
context of Sakharov's theory of induced gravity \cite{fro1},
\cite{fro2}. In Sakharov's proposal, the dynamical aspects of gravity
arise from the collective excitations of massive fields. Constraints
are then placed on these massive fields to cancel divergences and
ensure that the effective cosmological constant vanishes. Sakharov's
proposal is not expected to provide a fundamental description of
quantum gravity, but at scales below the Planck scale it may possess
features in common with other more fundamental descriptions. In common
with the entanglement entropy and thermal atmosphere approaches, black
hole entropy is explained as arising from the quantum field degrees of
freedom outside the black hole. However, in this case the formula for
black hole entropy involves a subtraction of the (divergent) mode
counting expression and an (equally divergent) expression for the
Noether charge operator, so that, in effect, only the massive fields
contribute to black hole entropy. The result of this subtraction
yields eq.(\ref{Sbh2}).

More recently, another approach to the calculation of black hole
entropy has been developed in the framework of quantum geometry
\cite{abck}, \cite{ak}. In this approach, if one considers a spacetime
containing an isolated horizon (see section \ref{cbht} above), the
classical symplectic form and classical Hamiltonian\footnote{The phase
space \cite{ack} considered here incorporates the isolated horizon
boundary conditions, i.e., only field variations that preserve the
isolated horizon structure are admitted.} each acquire an additional
boundary term arising from the isolated horizon \cite{afk}. These
additional terms are identical in form to that of a Chern-Simons
theory defined on the isolated horizon.  Classically, the fields on
the isolated horizon are determined by continuity from the fields in
the ``bulk'' and do not represent additional degrees of
freedom. However, in the quantum theory---where distributional fields
are allowed---these fields are interpreted as providing additional,
independent degrees of freedom associated with the isolated
horizon. One then counts the ``surface states'' of these fields on the
isolated horizon subject to a boundary condition relating the surface
states to ``volume states'' and subject to the condition that the area
of the isolated horizon (as determined by the volume state) lies
within a squared Planck length of the value $A$. This state counting
yields an entropy proportional to $A$ for black holes much larger than
the Planck scale. Unlike the entanglement entropy and thermal
atmosphere calculations, the state counting here yields finite results
and no cutoff need be introduced. However, the formula for entropy
contains a free parameter (the ``Immirzi parameter''), which arises
from an ambiguity in the loop quantization procedure, so the constant
of proportionality between $S$ and $A$ is not calculable.

The most quantitatively successful calculations of black hole entropy
to date are ones arising from string theory. It is believed that at
``low energies'', string theory should reduce to a 10-dimensional
supergravity theory (see \cite{m} for considerable further discussion
of the relationship between string theory and $10$-dimensional and
$11$-dimensional supergravity). If one treats this supergravity theory
as a classical theory involving a spacetime metric, $g_{ab}$, and
other classical fields, one can find solutions describing black
holes. On the other hand, one also can consider a ``weak coupling''
limit of string theory, wherein the states are treated perturbatively.
In the weak coupling limit, there is no literal notion of a black
hole, just as there is no notion of a black hole in linearized general
relativity. Nevertheless, certain weak coupling states can be
identified with certain black hole solutions of the low energy limit
of the theory by a correspondence of their energy and charges. (Here,
it is necessary to introduce ``D-branes'' into string perturbation
theory in order to obtain weak coupling states with the desired
charges.) Now, the weak coupling states are, in essence, ordinary
quantum dynamical degrees of freedom, so their entropy can be computed
by the usual methods statistical physics.  Remarkably, for certain
classes of extremal and nearly extremal black holes, the ordinary
entropy of the weak coupling states agrees exactly with the expression
for $A/4$ for the corresponding classical black hole states; see
\cite{ho} and \cite{peet} for reviews of these results. Recently, it
also has been shown \cite {cwm} that for certain black holes,
subleading corrections to the state counting formula for entropy
correspond to higher order string corrections to the effective
gravitational action, in precise agreement with eq.(\ref{Sbh}).

Since the formula for entropy has a nontrivial functional dependence
on energy and charges, it is hard to imagine that the agreement
between the ordinary entropy of the weak coupling states and black
hole entropy could be the result of a random coincidence. Furthermore,
for low energy scattering, the absorption/emission coefficients
(``gray body factors'') of the corresponding weak coupling states and
black holes also agree \cite{ms}. This suggests that there may be a
close physical association between the weak coupling states and black
holes, and that the dynamical degrees of freedom of the weak coupling
states are likely to at least be closely related to the dynamical
degrees of freedom responsible for black hole entropy. However, it
remains a challenge to understand in what sense the weak coupling
states could be giving an accurate picture of the local physics
occurring near (and within) the region classically described as a
black hole.

The relevant degrees of freedom responsible for entropy in the weak
coupling string theory models are associated with conformal field
theories.  Recently Carlip \cite{ca1}, \cite{ca2} has attempted to
obtain a direct relationship between the string theory state counting
results for black hole entropy and the classical Poisson bracket
algebra of general relativity. After imposing certain boundary
conditions correponding to the presence of a local Killing horizon,
Carlip chooses a particular subgroup of spacetime diffeomorphisms,
generated by vector fields $\xi^a$. The transformations on the phase
space of classical general relativity corresponding to these
diffeomorphisms are generated by Hamiltonians $H_\xi$. However, the
Poisson bracket algebra of these Hamiltonians is not isomorphic to the
Lie bracket algebra of the vector fields $\xi^a$ but rather
corresponds to a central extension of this algebra. A Virasoro algebra
is thereby obtained. Now, it is known that the asymptotic density of
states in a conformal field theory based upon a Virasoro algebra is
given by a universal expression (the ``Cardy formula'') that depends
only on the Virasoro algebra. For the Virasoro algebra obtained by
Carlip, the Cardy formula yields an entropy in agreement with
eq.(\ref{Sbh2}). Since the Hamiltonians, $H_\xi$, are closely related
to the corresponding Noether currents and charges occurring in the
derivation of eqs.(\ref{Sbh}) and (\ref{Sbh2}), Carlip's approach
holds out the possibility of providing a direct, general explanation
of the remarkable agreement between the string theory state counting
results and the classical formula for the entropy of a black hole.

\newpage
\section{Open Issues}
\label{oi}

The results described in the previous sections provide a remarkably
compelling case that stationary black holes are localized thermal
equilibrium states of the quantum gravitational field, and that the
laws of black hole mechanics are simply the ordinary laws of
thermodynamics applied to a system containing a black hole. Although
no results on black hole thermodynamics have been subject to any
experimental or observational tests, the theoretical foundation of
black hole thermodynamics appears to be sufficiently firm as to provide a
solid basis for further research and speculation on the nature of
quantum gravitational phenomena. In this section, I will briefly
discuss two key unresolved issues in black hole thermodynamics may
shed considerable further light upon quantum gravitational physics.

\subsection{Does a Pure Quantum State Evolve to a Mixed State in the 
Process of Black Hole Formation and Evaporation?}
\label{pm}

In classical general relativity, the matter responsible for the
formation of a black hole propagates into a singularity lying within
the deep interior of the black hole. Suppose that the matter which
forms a black hole possesses quantum correlations with matter that
remains far outside of the black hole. Then it is hard to imagine how
these correlations could be restored during the process of black hole
evaporation unless gross violations of causality occur. In fact, the
semiclassical analyses of the Hawking process show that, on the
contrary, correlations between the exterior and interior of the black
hole are continually built up as it evaporates (see \cite{w4} for
further discussion). Indeed, these correlations play an essential role
in giving the Hawking radiation an exactly thermal character
\cite{w7}.

As already mentioned in subsection \ref{av} above, an isolated black
hole will ``evaporate'' completely via the Hawking process within a
finite time. If the correlations between the inside and outside of the
black hole are not restored during the evaporation process, then by
the time that the black hole has evaporated completely, an initial
pure state will have evolved to a mixed state, i.e., ``information''
will have been lost. In a semiclassical analysis of the evaporation
process, such information loss does occur and is ascribable to the
propagation of the quantum correlations into the singularity within
the black hole. A key unresolved issue in black hole thermodynamics is
whether this conclusion continues to hold in a complete quantum theory
of gravity. On one hand, arguments can be given \cite{w4} that
alternatives to information loss---such as the formation of a high
entropy ``remnant'' or the gradual restoration of correlations during
the late stages of the evaporation process---seem highly
implausible. On the other hand, it is commonly asserted that the
evolution of an initial pure state to a final mixed state is
in conflict with quantum mechanics. For this reason, the issue of
whether a pure state can evolve to a mixed state in the process of
black hole formation and evaporation is usually referred to as the
``{\it black hole information paradox}''.

There appear to be two logically independent grounds for the claim
that the evolution of an initial pure state to a final mixed state is
in conflict with quantum mechanics: (1) Such evolution is asserted to
be incompatible with the fundamental principles of quantum theory,
which postulates a unitary time evolution of a state vector in a
Hilbert space. (2) Such evolution necessarily gives rise to violations
of causality and/or energy-momentum conservation and, if it occurred
in the black hole formation and evaporation process, there would be
large violations of causality and/or energy-momentum (via processes
involving ``virtual black holes'') in ordinary laboratory physics.

With regard to (1), within the semiclassical framework, the evolution
of an initial pure state to a final mixed state in the process of
black hole formation and evaporation can be attributed to the fact
that the final time slice fails to be a Cauchy surface for the
spacetime \cite{w4}. No violation of any of the local laws of quantum
field theory occurs. In fact, a closely analogous evolution of an
initial pure state to a final mixed state occurs for a free, massless
field in Minkowski spacetime if one chooses the final ``time'' to be a
hyperboloid rather than a hyperplane \cite{w4}. (Here, the
``information loss'' occurring during the time evolution results from
radiation to infinity rather than into a black hole.) Indeed, the
evolution of an initial pure state to a final mixed state is naturally
accomodated within the framework of the algebraic approach to quantum
theory \cite{w4} as well as in the framework of generalized quantum
theory \cite{ha}.

The main arguments for (2) were given in \cite{bps} (see also
\cite{ehns}). However, these arguments assume that the effective
evolution law governing laboratory physics has a ``Markovian''
character, so that it is purely local in time. As pointed out in
\cite{uw3}, one would expect a black hole to retain a ``memory''
(stored in its external gravitational field) of its energy-momentum,
so it is far from clear that an effective evolution law modeling the
process of black hole formation and evaporation should be Markovian in
nature. Furthermore, even within the Markovian context, it is not
difficult to construct models where rapid information loss occurs at
the Planck scale, but negligible deviations from ordinary dynamics
occur at laboratory scales \cite{uw3}.

For the above reasons, I do not feel that the issue of whether a pure
state evolves to a mixed state in the process of black hole formation
and evaporation should be referred to as a ``paradox''. Nevertheless,
the resolution of this issue is of great importance: If pure states
remain pure, then our basic understanding of black holes in classical
and semiclassical gravity will have to undergo significant revision in
quantum gravity. On the other hand, if pure states evolve to mixed
states in a fully quantum treatment of the gravitational field, then
at least the aspect of the classical singularity as a place where
``information can get lost'' must continue to remain present in
quantum gravity. In that case, rather than ``smooth out'' the
singularities of classical general relativity, one might expect
singularities to play a fundamental role in the formulation of quantum
gravity \cite{p}. Thus, the resolution of this issue would tell us a
great deal about both the nature of black holes and the existence of
singularities in quantum gravity.

\subsection{What (and Where) are the Degrees of Freedom Responsible for 
Black Hole Entropy?}
\label{dof}

The calculations described in section \ref{bhe} yield a seemingly
contradictory picture of the degrees of freedom responsible for black
hole entropy. In the entanglement entropy and thermal atmosphere
approaches, the relevant degrees of freedom are those associated with
the ordinary degrees of freedom of quantum fields outside of the black
hole. However, the dominant contribution to these degrees of freedom
comes from (nearly) Planck scale modes localized to (nearly) a Planck
length of the black hole, so, effectively, the relevant degrees of
freedom are associated with the horizon. In the quantum geometry
approach, the relevant degrees of freedom are also associated with the
horizon but appear to have a different character in that they reside
directly on the horizon (although they are constrained by the exterior
state). Finally the string theory calculations involve weak coupling
states, so it is not clear what the the degrees of
freedom of these weak coupling states would correspond to in a low
energy limit where these states may admit a black hole
interpretation. However, there is no indication in the calculations
that these degrees of freedom should be viewed as being localized near
the black hole horizon.

The above calculations are not necessarily in conflict with each
other, since it is possible that they each could represent a
complementary aspect of the same physical degrees of freedom.
Nevertheless, it seems far from clear as to whether we should think of
these degrees of freedom as residing outside of the black hole (e.g.,
in the thermal atmosphere), on the horizon (e.g., in Chern-Simons
states), or inside the black hole (e.g., in degrees of freedom
associated with what classically corresponds to the singularity deep
within the black hole). 

The following puzzle \cite{w8} may help bring into focus some of the
issues related to the degrees of freedom responsible for black hole
entropy and, indeed, the meaning of entropy in quantum gravitational
physics. As we have already discussed, one proposal for accounting for
black hole entropy is to attribute it to the ordinary entropy of its
thermal atmosphere. If one does so, then, as previously mentioned in
section \ref{bhe} above, one has the major puzzle of explaining why
the quantum field degrees of freedom near the horizon contribute
enormously to entropy, whereas the similar degrees of freedom that are
present throughout the universe---and are locally indistinguishable
from the thermal atmosphere---are treated as mere ``vacuum
fluctuations'' which do not contribute to entropy. But perhaps an even
greater puzzle arises if we assign a negligible entropy to the thermal
atmosphere (as compared with the black hole area, $A$), as would be
necessary if we wished to attribute black hole entropy to other
degrees of freedom.  Consider a black hole enclosed in a reflecting
cavity which has come to equilibrium with its Hawking
radiation. Surely, far from the black hole, the thermal atmosphere in
the cavity must contribute an entropy given by the usual formula for a
thermal gas in (nearly) flat spacetime. However, if the thermal
atmosphere is to contribute a negligible total entropy (as compared
with $A$), then at some proper distance $D$ from the horizon much
greater than the Planck length, the thermal atmosphere must contribute
to the entropy an amount that is much less than the usual result
($\propto T^3$) that would be obtained by a naive counting of
modes. If that is the case, then consider a box of ordinary thermal
matter at infinity whose energy is chosen so that its floating point
would be less than this distance $D$ from the horizon.  Let us now
slowly lower the box to its floating point. By the time it reaches its
floating point, the contents of the box are indistinguishable from the
thermal atmosphere, so the entropy within the box also must be less
than what would be obtained by usual mode counting arguments. It
follows that the entropy within the box must have decreased during the
lowering process, despite the fact that an observer inside the box
still sees it filled with thermal radiation and would view the
lowering process as having been adiabatic. Furthermore, suppose one
lowers (or, more accurately, pushes) an empty box to the same distance
from the black hole. The entropy difference between the empty box and
the box filled with radiation should still be given by the usual mode
counting formulas. Therefore, the empty box would have to be assigned
a negative entropy.

I believe that in order to gain a better understanding of the degrees
of freedom responsible for black hole entropy, it will necessary to
achieve a deeper understanding of the notion of entropy itself. Even
in flat spacetime, there is far from universal agreement as to the
meaning of entropy---particularly in quantum theory---and as to the
nature of the second law of thermodynamics. The situation in general
relativity is considerably murkier \cite{w2}, as, for example, there
is no unique, rigid notion of ``time translations'' and classical
general relativistic dynamics appears to be incompatible with any
notion of ``ergodicity''.  It seems likely that a new conceptual
framework will be required in order to have a proper understanding of
entropy in quantum gravitational physics.

\section{Acknowledgements}
\label{acknowledgements}

This research was supported in part by NSF grant PHY 95-14726 to the
University of Chicago.

\end{document}